\documentclass[12pt]{article}
\usepackage{epsfig}
\usepackage{epsfig}

\newcommand{\be}{\begin{equation}}
\newcommand{\ee}{\end{equation}}
\newcommand{\ba}{\begin{eqnarray}}
\newcommand{\ea}{\end{eqnarray}}

\newcommand{\ed}{\end{document}}
\newcommand{\baz}{\begin{eqnarray*}}
\newcommand{\eaz}{\end{eqnarray*}}
\newcommand{\bb}{\begin{thebibliography}}
\newcommand{\eb}{\end{thebibliography}}

\begin{document}
\begin{titlepage}

\begin{center}

\vspace{5cm}

{\Large{\bf Instantons and Spin-Flavor effects in Hadron Physics}}
\vspace{0.50cm}\\
 N.I. Kochelev\footnote{kochelev@theor.jinr.ru}\\
\vskip 1ex { \it Bogoliubov Laboratory of Theoretical Physics,
Joint Institute for Nuclear Research, Dubna, Moscow region,
141980, Russia}
\end{center}

\vskip 0.5cm \centerline{\bf Abstract}

We discuss the role of  instantons in the spectroscopy
of  ordinary and exotic hadrons  as well as in  high energy
 reactions. We argue that the  instanton
induced flavor- and spin-dependent quark-quark and quark-gluon interactions
can explain many features of the  hadron spectrum.
The observed anomalous spin and flavor effects in  various
reactions with hadrons
can also be understood within the instanton model for QCD vacuum.

\vskip 0.3cm \leftline{Pacs: 12.38.Aw, 12.38.Lg, 12.39.Ba, 12.39.x}
\leftline{Keywords: quarks,
instanton,multiquark, high energy scattering}

\end{titlepage}

\vspace{1cm}

\setcounter{footnote}{0}

\section{ Introduction}
The existence of instanton, a
strong nonperturbative fluctuation of gluon fields, in the QCD vacuum
is considered as a primary factor of chiral
  and $U(1)_A$ symmetry violations
 (see reviews
\cite{shuryak, diakonov}). A well-known example of an instanton induced
 interaction is the
famous t'Hooft quark-quark interaction, which can be obtained from the
consideration of
the so-called quark zero-modes in the instanton field \cite{thooft}.
Another example for the instanton induced chirality-flip
interaction is the non-perturbative quark-gluon chromomagnetic
interaction \cite{kochelev2, diakonov}.
In this Letter we discuss the effects of these interactions in the
 hadron spectroscopy
and reactions with
hadrons. Our main purpose is to stress the importance of the
instanton induced quark-quark and quark-gluon
interactions in the spin and flavor structure of hadrons and
in high energy reactions in a few GeV region for momentum transfer.

\section { Instanton induced interaction and structure
\\ of ordinary and exotic
hadrons}

Recently, the evidence for the existence of the exotic $\Theta^+$ pentaquark state
with the strangeness $S=+1$ has been obtained. In spite of the unclear
experimental status of this state at the present time,
the fundamental question about the existence of   bound multiquarks
has not receive a certain answer so far. This is the reason why
 the search for quark exotic states
is included in many experimental programs at current and planned
facilities.  Such experimental activity calls for the reconsideration of
old
theoretical approaches to the multiquark spectroscopy which was  based on the
assumption of the dominance of a perturbative one-gluon exchange between
quarks  inside the bag
 \cite{jaffem}.  Within such an approach, the correlations between
quarks in the bag are very weak and, therefore, one expects
that multiquarks should have rather large masses and widths.
Furthermore, the bag model predicts a very large number
of these states which should mix with each other as well as with the colorless hadronic
resonances carrying the same quantum numbers.

The first hint at the possibility to have
a light pentaquark with a small width was found within  the soliton model for baryons
\cite{diakonov1}. In this model the peculiarities in
the structure of pentaquarks
 are determinated by the collective dynamics of quarks in the background of
the meson field. The explanation of a small mass and width of the pentaquark
 was also  given
within  the constituent quark model
 based on the possible cluster structure of the pentaquark arising from
the attraction in some diquark state due to
 the perturbative one-gluon exchange (OGE)
 between quarks
\cite{Jaffe:2003sg}.

In the alternative approach to the hadron spectroscopy, in which
nonperturbative, instanton induced interaction between constituent
 quarks plays dominant role,
was developed in \cite{kochelev1}
(see a recent review \cite{Kochelev:2005xn}). In this model, many
features of the observed spectrum of ordinary hadrons can be
 described by the contribution
arised from
the effective two-body and three-body t'Hooft interactions:
\begin{eqnarray}
{\cal H}_{eff}^{(2)}(r)&=&-V_2\sum_{i\neq j}\frac{1}
{m_im_j}\bar q_{iR}(r)q_{iL}(r) \bar
q_{jR}(r)q_{jL}(r)
\bigg[1+\frac{3}{32}(\lambda_i^a\lambda_j^a)\nonumber\\
&&+\frac{9}{32}(\vec{\sigma_i}\cdot\vec{\sigma_j}\lambda_i^a\lambda_j^a)\bigg]
+(R\longleftrightarrow L), \label{thooft2}
\end{eqnarray}
where $m_i=m_i^{cur} +m^*$ is the effective quark mass in the
nonperturbative vacuum.
In particular, such an interaction helps
solving  the famous $U(1)_A$ problem, which is related to the large mass
of $\eta^\prime $ meson
and, simultaneously,  produces a very light $\pi$ meson state. It is well known
that it is extremely difficult to obtain a heavy $\eta^\prime$ and a
light $\pi$ meson within the OGE model.
Furthermore, the instanton based constituent quark model
has been used  to calculate the properties of various tetraquark states
 and to study the  structure of the $2\Lambda$, so-called H-dibaryon, state.
In comparison with the OGE models quite a different spectrum
 of mass of multiquarks
was obtained
\cite{DKZ}, \cite{H}.

For multiquark hadrons with open and hidden strangeness
  and for the reactions
 including  the strange quark
the three-body t'Hooft interaction
\begin{eqnarray}
{\cal H}_{eff}^{(3)}(r)&=&- V_3\prod_{i=u,d,s}\bar
q_{iR}(r)q_{iL}(r) \bigg[1+\frac{3}{32}(\lambda_u^a\lambda_d^a+{\rm perm.})
\nonumber\\
&&+\frac{9}{32}(\vec{\sigma_u}\cdot\vec{\sigma_d}\lambda_u^a\lambda_d^a+{\rm perm.})
 -\frac{9}{320}d^{abc}\lambda^a\lambda^b\lambda^c
(1-3(\vec{\sigma_u}\cdot\vec{\sigma_d}+{\rm perm.}))\nonumber\\
&&-\frac{9f^{abc}}{64}\lambda^a\lambda^b\lambda^c
(\vec{\sigma_u}\times\vec{\sigma_d})\cdot\vec{\sigma_s}\bigg]+(R\longleftrightarrow
L), \label{thooft4}
\end{eqnarray}
 might also be important.

Recently, this model has been applied to the pentaquark
 spectroscopy  \cite{KLV,Lee:2004dp}.
It was argued that a specific flavor- and spin-dependent  t'Hooft multiquark interaction
forms a certain type of two- and three-particle clusters inside the multiquark
hadron.
As the result, the
bound multiquark states might appear in the instanton field.
The importance of the instantons in the multiquark dynamics was confirmed
by direct calculation of the light pentaquark and tetraquark masses
within the QCD sum rules in \cite{Lee:2005ny,Lee:2006vk,Lee:2007mva}.

We should mention   that the possibility of the scalar ud-diquark formation
inside the nucleon due to the instanton interaction was also discussed  in
\cite{Schafer:1994fd} and within the  QCD sum rule approach (QCDSR)
  in \cite{Dorokhov:1989zw,SSV}.
Furthermore,  the QCDSR  calculation  carried out in \cite{Lee:2004dp}
confirms the conclusion of the constituent model \cite{KLV}
about the appearance of the light $ud\bar s$ triquark state in the instanton field.
Instantons also play a very important role in glueball physics
\cite{Schafer:1994fd,forkel}
 and, in particular,
are responsible for mass splitting of the parity partners in the glueball sector
\cite{Kochelev:2005vd,Mathieu:2008bf}.
Finally, we should emphasize that the instanton induced multiquark interaction
also gives rise to some
weak hadronic decays and, particularly, can be considered
as a fundamental QCD  mechanism for the
empirical $\Delta I=1/2$ rule found
in the weak $\Delta S=1$ decays  \cite{Kochelev:2001pp}.

\section { Spin and flavor structure of nucleon}

More than fifteen years ago we argued that an instanton induced spin-flip
 interaction should lead to negative polarization of sea quarks inside polarized nucleon and
 to valence quark depolarization \cite{Dorokhov:1993ym,Kochelev:1997ux}.
In this type of approach, it is not necessary to have a sizeable gluon polarization
to explain
the famous "spin crises" \cite{Anselmino:1994gn}. The recent results
 obtained by the STAR
\cite{Gagliardi:2008qw}
and COMPASS Collaborations \cite{Stolarski:2008jc} give a  small value of gluon
polarization and, therefore, confirm our prediction.
Furthermore, we   showed  \cite{Dorokhov:1993ym,Dorokhov:1993fc}
 that due to the Pauli principle
for quarks in the instanton field,
large sea quark polarization should also be accommodated
by large flavor asymmetry in the proton sea. Indeed, this
$\bar u-\bar d$ asymmetry was found
in the Drell-Yan muon pair  production from analyses of the
  cross section of pp and pn scatterings
 \cite{Webb:2003ps,Towell:2001nh}.

\section { Instanton effects in high energy reactions}

The significant single-spin asymmetries (SSA) in meson production in
semi-inclusive deep-inelastic scattering (SIDIS) observed by the HERMES
Collaboration at DESY \cite{HERMES, HERMESK} is the challenge
for the pQCD approach to spin effects in  strong interactions. One of the
unexpected phenomena found by HERMES is the large Sivers asymmetry
for the $ K^+$ meson. Such asymmetry is in contradiction with expectations
of the naive pQCD based on the picture in which the main contribution to
$K^+$ SSA comes from u-valence quark fragmentation \cite{efremov}.
Recently, a new approach to the SSA in SIDIS has been suggested \cite{SSA}.
It is based on the instanton induced final state arising from
multiquark interaction (Eq.\ref{thooft4}).

\begin{figure}[h]
\begin{minipage}[c]{8cm}
\vspace*{1.5cm}
\centerline{\epsfig{file=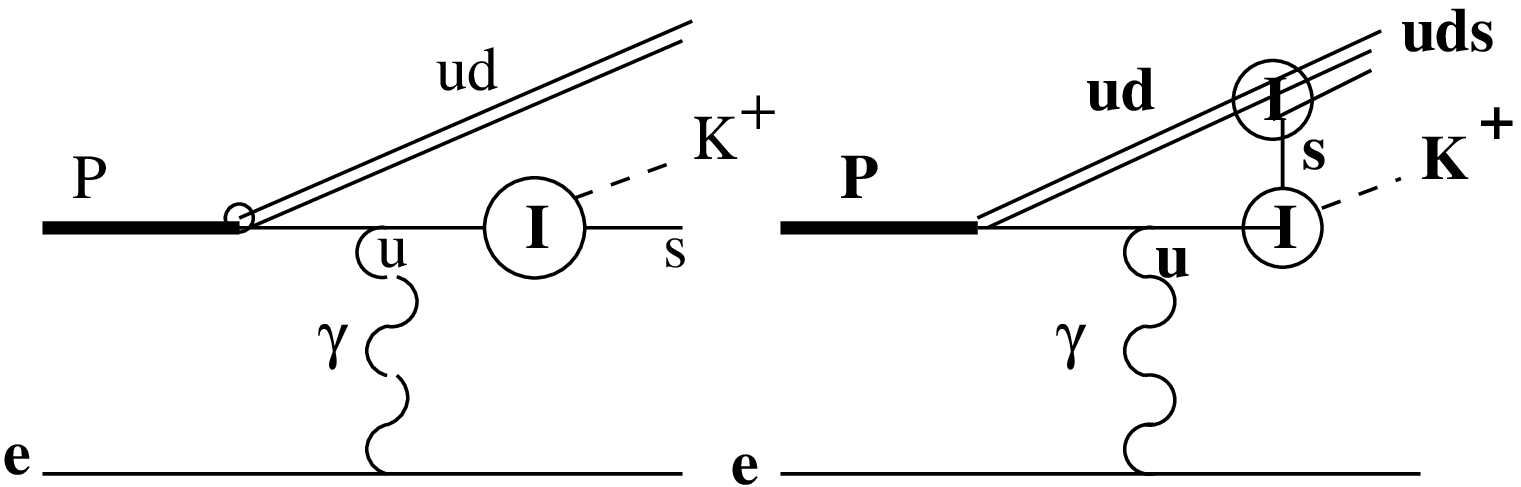,width=9cm,height=3cm, angle=0}}\
 \caption{
The diagrams contributing to $K^+$ SSA. The symbol I denotes the instanton
(antiinstanton).}
\end{minipage}
\hspace*{0.5cm}
\vspace*{-0.5cm}
\begin{minipage}[c]{8cm}
\centering \centerline{\epsfig{file=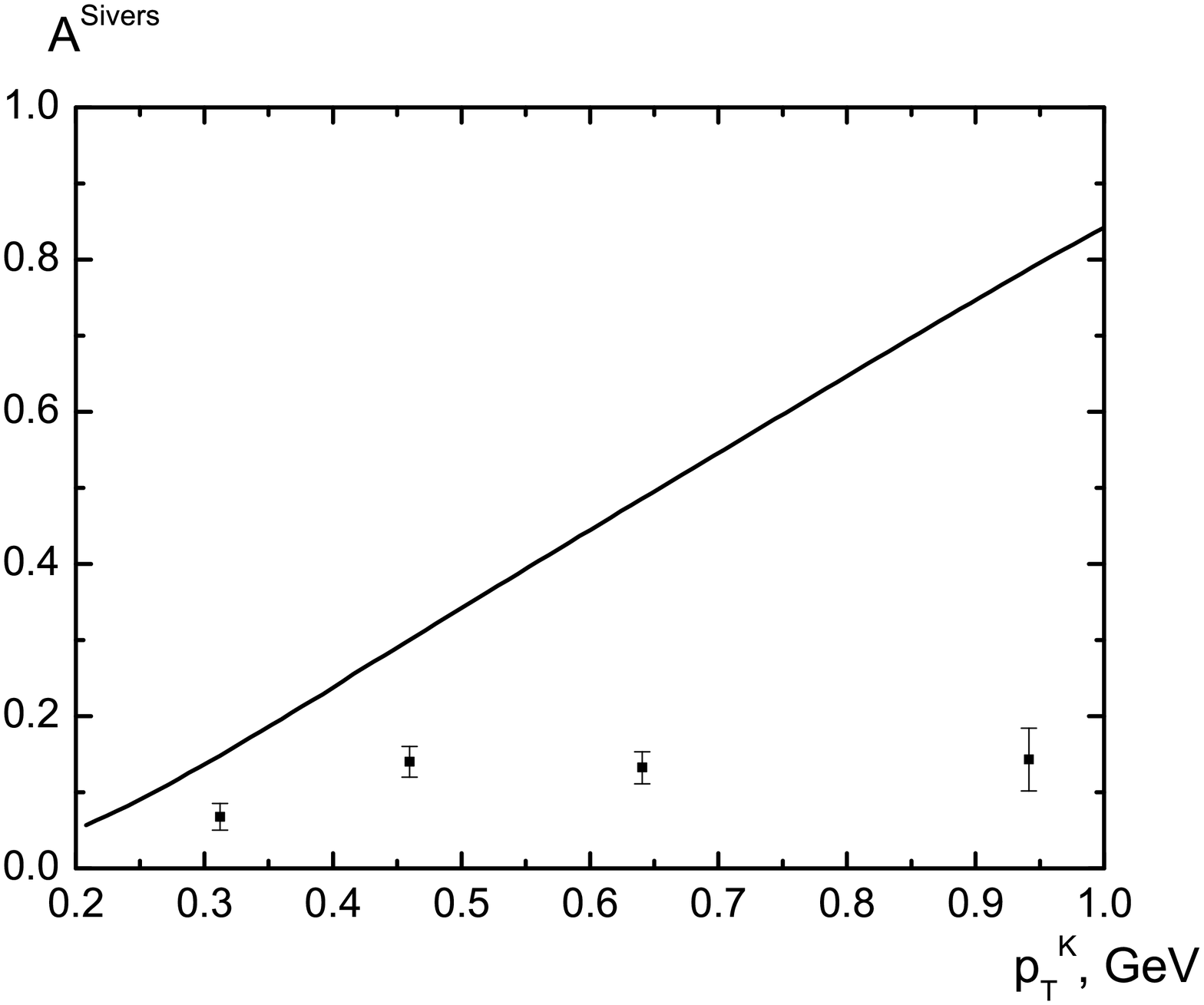,width=8cm,angle=0}}\
\vspace*{-0.5cm}
 \caption{ The  dependence of $K^+$ Sivers
asymmetry on $p_{\bot}$ in the comparison with  preliminary
HERMES data \cite{HERMESK}.}
\end{minipage}
\end{figure}

Some of the diagrams responsible for $K^+$ SSA in
SIDIS are shown in Fig.1. The result
presented in Fig.2  shows that, indeed,
the interference of such types of
diagrams may give the large $K^+$-meson Sivers asymmetry.
For more detailed comparison with the data one should take into account
form factors in the nonperturbative s-(ud) and s-u vertices.

\begin{figure}[h]
\begin{minipage}[c]{8cm}
\vspace*{1.2cm}
\centerline{\epsfig{file=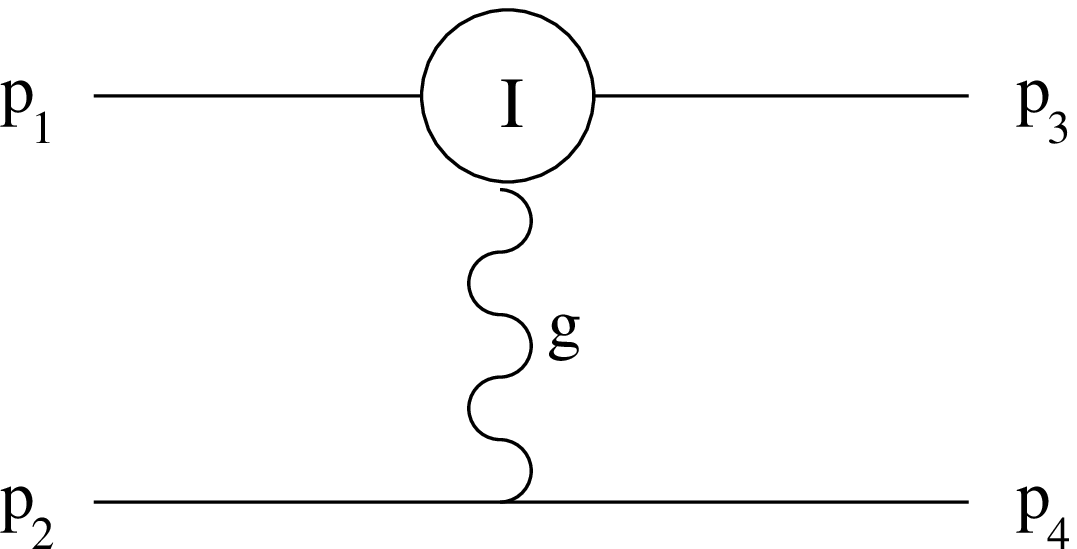,width=6cm,angle=0}}\
 \caption{
The quark chromomagnetic  moment contribution to the high energy
quark-quark scattering.}
\end{minipage}
\hspace*{0.5cm}
%\vspace*{-1.5cm}
\begin{minipage}[c]{8cm}
\centering \centerline{\epsfig{file=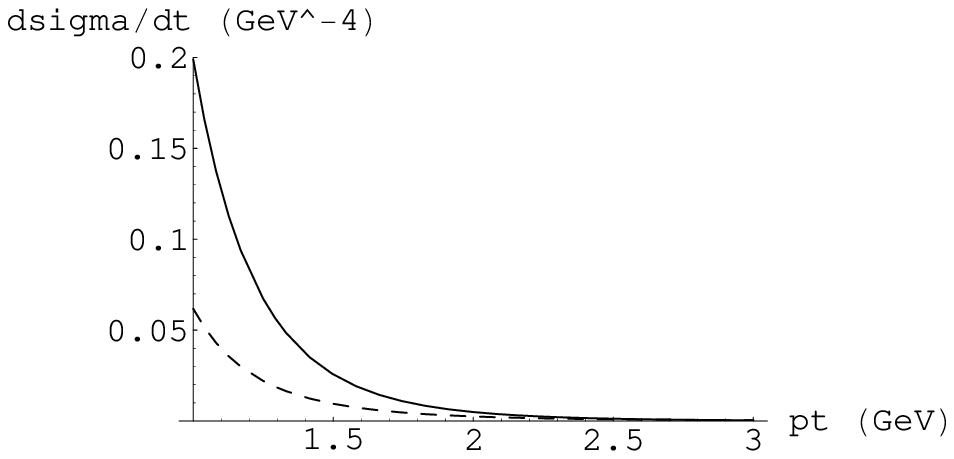,width=8cm,angle=0}}\
\vspace*{-0.5cm}
\caption{Perturbative  (dashed line)  and nonperturbative (solid
line) quark-quark cross sections as  functions of transverse
momentum.}
\end{minipage}
\end{figure}
Instantons can also contribute to the quark-quark, quark-gluon and gluon-gluon scattering
at high energy. It is well known that the t-channel gluon exchange  leads to
a nonzero contribution in the high energy partonic cross section.
There are two possible contributions to gluonic exchange arising from instantons.
One of them
 contributes to the gluon propagator \cite{Dorokhov:2004fb} and
determines its infrared behavior. The other is related to
 the nonperturbative correction to the quark-gluon vertex
 and can be treated
as a quark chromomagnetic moment induced by instantons
\cite{kochelev2, diakonov}
\begin{equation}
{\cal L_I}^{chromo}= -i\frac{g_s\mu_a}{2m^*_i} \bar q
\sigma_{\mu\nu}t^aG^a_{\mu\nu}q,
 \label{chrom}
 \end{equation}
  where $ \mu_a$ is the quark anomalous
chromomagnetic moment,  $G_{\mu\nu}$ is the gluon field strength.
For the off-shell gluon with virtuality $q$ the instanton induced
quark-gluon vertex, Eq. (\ref{chrom}), should be multiplied by the
instanton form factor
 \begin{equation}
F(z)=\frac{4}{z^2}-2K_2(z), \label{form2}
 \end{equation} where
$z=q\rho_c/2$ and $\rho_c$ is the average size of instanton
in the QCD vacuum. The value of the quark anomalous chromomagnetic
moment is proportional to  the instanton packing  fraction
$f=n_c\pi^2\rho_c^4\approx 0.1$ in the QCD vacuum, where $n_c$ is
the instanton density. A remarkable property of this new
quark-gluon interaction is its chirality structure. Indeed, it leads to
 spin-flip of quark and, therefore, it might be responsible for
large spin asymmetries observed in high energy reactions
\cite{kochelev2}.
Recently,  it was shown that the contribution of this interaction
to the quark-quark high energy scattering cross section
arising from the diagram presented in Fig.1
 exceeds  the pQCD one-gluon exchange
 contribution in  the  region of transverse
momentum $p_\bot < 3$ GeV.  So one can expect the appearance
of the perturbative QCD regime only  at a large enough value of
$p_\bot$.

\section{Conclusion}

 We discussed the effects of the nonperturbative
structure of the  QCD vacuum in the strong interaction. It is shown that
specific instanton induced interactions between hadron constituents
play a very important role in the spectroscopy of ordinary and
exotic hadrons as well as in
high energy reactions.

\section{Acknowledgment}

The author is grateful to  Alexander Dorokhov, Hee-Jung Lee, Dong-Pil Min and
 Vicente Vento for the fruitfull collaboration. This work was
supported by Belarus-JINR grant.


\vspace{0cm}

\begin{thebibliography}{99}

\bibitem{shuryak} T. Sch\"afer and E.V. Shuryak,
Rev. Mod. Phys. {\bf 70} (1998) 1323.

\bibitem{diakonov} D. Diakonov, Prog. Par. Nucl. Phys.
{\bf 51} (2003) 173.

\bibitem{thooft}
   G.~'t Hooft,
   Phys.\ Rev.\ {\bf D14} (1976) 3432.

\bibitem{kochelev2}
   N.~I.~Kochelev,
   Phys.\ Lett.\  {\bf B426} (1998) 149.

\bibitem{jaffem} R.L. Jaffe, Phys. Rev. {\bf D15} (1977) 267; 281.

\bibitem{diakonov1} D. Diakonov, V. Petrov and M. Polyakov, Z.
Phys. {\bf A359} (1997) 305.

\bibitem{Jaffe:2003sg}
  R.~L.~Jaffe and F.~Wilczek,
  Phys.\ Rev.\ Lett.\  {\bf 91} (2003) 232003.



 \bibitem{kochelev1}
  N.~I.~Kochelev,
  Sov.\ J.\ Nucl.\ Phys.\  {\bf 41} (1985) 291
  [Yad.\ Fiz.\  {\bf 41} (1985) 456];
A.~E.~Dorokhov, Yu.~A.~Zubov and N.~I.~Kochelev,
  Sov.\ J.\ Part.\ Nucl.\  {\bf 23} (1992) 522 and references therein.

\bibitem{Kochelev:2005xn}
  N.~I.~Kochelev,
  %``QCD vacuum structure and hadron properties,''
  Phys.\ Part.\ Nucl.\  {\bf 36} (2005) 608
  [Fiz.\ Elem.\ Chast.\ Atom.\ Yadra {\bf 36} (2005) 1157].

\bibitem{DKZ}A.E. Dorokhov, N.I.Kochelev, Yu.A. Zubov,
Yad. Fiz. {\bf 50} (1989) 1717;
 Z. Phys. {\bf C65} (1995) 667.

\bibitem{H}
A.E. Dorokhov and N.I.Kochelev, Preprint JINR-E2-86-847 (1986),
hep-ph/0411362;   N.I. Kochelev, JETP
Lett. {\bf 70} (1999) 491.


\bibitem{KLV} N.I.Kochelev, H.-J. Lee and V. Vento, Phys. Lett.
{\bf B594} (2004) 87.

\bibitem{Lee:2004dp}
  H.~J.~Lee, N.~I.~Kochelev and V.~Vento,
  Phys.\ Lett.\   {\bf B610} (2005) 50.

\bibitem{Lee:2005ny}
  H.~J.~Lee, N.~I.~Kochelev and V.~Vento,
  Phys.\ Rev.\  D {\bf 73} (2006) 014010

\bibitem{Lee:2006vk}
H.~J.~Lee and N.~I.~Kochelev,
  Phys.\ Lett.\   {\bf B642} (2006) 358

\bibitem{Lee:2007mva}
  H.~J.~Lee and N.~I.~Kochelev,
  arXiv:hep-ph/0702225 (to be published in Phys. Rev. D).


\bibitem{SSV} T. Sch\"afer, E.V. Shuryak and J.J.M. Verbaarschot,
 Nucl. Phys. {\bf B412} (1994) 143.

\bibitem{Dorokhov:1989zw}
  A.~E.~Dorokhov and N.~I.~Kochelev,
  Z.\ Phys.\   {\bf C46} (1990) 281.



\bibitem{Kochelev:2001pp}
  N.~I.~Kochelev and V.~Vento,
  Phys.\ Rev.\ Lett.\  {\bf 87} (2001) 111601.






\bibitem{Schafer:1994fd}
  T.~Schafer and E.~V.~Shuryak,
   Phys.\ Rev.\ Lett.\  {\bf 75}(1995) 1707;
  M.~C.~Tichy and P.~Faccioli,
  arXiv:0711.3829 [hep-ph].



\bibitem{forkel}
  H.~Forkel,
  Phys.\ Rev.\   {\bf D71}(2005) 054008;
  A.~l.~Zhang and T.~G.~Steele, Nucl.\ Phys.\   {\bf A728} (2003) 165.


\bibitem{Kochelev:2005vd}
  N.~Kochelev and D.~P.~Min,
  Phys.\ Lett.\   {\bf B633} (2006) 283.


\bibitem{Mathieu:2008bf}
  V.~Mathieu, F.~Buisseret and C.~Semay,
  %``Gluons in glueballs: Spin or helicity?,''
  Phys.\ Rev.\  D {\bf 77} (2008) 114022
  [arXiv:0802.0088 [hep-ph]].


\bibitem{Dorokhov:1993ym}
  A.~E.~Dorokhov, N.~I.~Kochelev and Yu.~A.~Zubov,
  Int.\ J.\ Mod.\ Phys.\  A {\bf 8} (1993) 603.


\bibitem{Kochelev:1997ux}
  N.~I.~Kochelev,
  Phys.\ Rev.\   {\bf D57} (1998) 5539.





\bibitem{Anselmino:1994gn}
  M.~Anselmino, A.~Efremov and E.~Leader,
  %``The theory and phenomenology of polarized deep inelastic scattering,''
  Phys.\ Rept.\  {\bf 261} (1995) 1


\bibitem{Gagliardi:2008qw}
  C.~A.~Gagliardi  [STAR Collaboration], DIS2008,
  %``Jet Production in Polarized pp Collisions at RHIC,''
  arXiv:0808.0858 [hep-ex].

\bibitem{Stolarski:2008jc}
  M.~Stolarski, on behalf of COMPASS Collaboration,  DIS2008,
  %``Measurement of Delta G/G from high transverse momentum hadron pairs in
  %COMPASS,''
  arXiv:0809.1803 [hep-ex].


\bibitem{Dorokhov:1993fc}
  A.~E.~Dorokhov and N.~I.~Kochelev,
  %``Instanton induced asymmetric quark configurations in the nucleon and parton
  %sum rules,''
  Phys.\ Lett.\  B {\bf 304}, 167 (1993).



\bibitem{Webb:2003ps}
  J.~C.~Webb {\it et al.}  [NuSea Collaboration],
  arXiv:hep-ex/0302019.

\bibitem{Towell:2001nh}
  R.~S.~Towell {\it et al.}  [FNAL E866/NuSea Collaboration],
  Phys.\ Rev.\   {\bf D64} (2001) 052002.

\bibitem{HERMES} HERMES Collaboration, A. Airapetian, et al., Phys. Rev. Lett. {\bf 84} (2000) 4047;
Phys. Rev. {\bf D64} (2001) 097101; Phys. Lett. {\bf B622} (2005) 14.

\bibitem{HERMESK} V.I. Korotkov, on behalf of the HERMES
Collaboration, ICHEP-06, July 2006, Moscow.

\bibitem{efremov}
A.~V.~Efremov,
  Annalen Phys.\  {\bf 13} (2004) 651.


\bibitem{SSA}
N.I. Kochelev, talk presented at XIII International Conference
"Selected Problems of Modern Theoretical Physics", June 23-27,
2008, Dubna, Russia; N.I. Kochelev,
Yu.M. Bystritskiy,
E.A. Kuraev,
W.-D. Nowak, to be published.


\bibitem{Dorokhov:2004fb}
  A.~E.~Dorokhov and I.~O.~Cherednikov,
  %``Instanton effects in quark form factor and quark quark scattering at  high
  %energy,''
  Annals Phys.\  {\bf 314} (2004) 321.


\bibitem{Kochelev:2006ny}
  N.~Kochelev,
  %``Soft contribution to quark-quark scattering induced by an anomalous
  %chromomagnetic interaction,''
  JETP Lett.\  {\bf 83} (2006) 527.






\end{thebibliography}
\end{document}